\documentclass{ws-procs9x6}
\usepackage{epsfig}
\def\tmi{T$_{mi}$}
\def\tc{T$_{c}$}

\def\sb#1{$_{#1}$}
\def\sp#1{$^{#1}$}

\def\lsco{La$_{2-x}$Sr$_x$CuO$_4$}

\def\etal{{\it et al.}}
\def\lcmo{La$_{1-x}$Ca$_{x}$MnO$_{3}$}

\begin{document}

\title{Strain, nano-phase separation, multi-scale structures and function of advanced materials}

\author{S. J. L. Billinge}

\address{Dept. Physics and Astronomy \\
Michigan State University\\
4263 Biomedical Physical Sciences Building, \\ 
East Lansing, MI 48824, USA\\ 
E-mail: billinge@pa.msu.edu}


\maketitle

\abstracts{
Recent atomic pair distribution function results from our group from manganites and
cuprate systems are reviewed in light of the presence of multi-scale structures.  These
structures have a profound effect on the material properties.
}

\section{Introduction}
In advanced materials with interesting functionality a realization is growing that
the properties depend sensitively on complex structures on different length-scales from
atomic to macroscopic.\cite{salje;cp00,sheno;prb99,stojk;prb00,rasmu;prl01,castr;prb01}  
The challenge to experimentalists is properly to characterize
such materials.  This will be a prerequisite to obtaining a complete theoretical
understanding of these materials.  It is particularly difficult because many of the
existing technologies for studying a material's structure quantitatively reveal only
average properties such as the crystal structure.  However, key information about higher 
level structures, potentially driving the interesting properties, is contained in deviations
from this average picture.  New approaches to studying structure are therefore required to
solve this problem.  A number of imaging techniques now exist such as TEM, STM, AFM and so on
and these are proving to be extremely important.\cite{uehar;n99,pan;n01}  What is needed in
addition is bulk probes that provide quantitative information about atomic structures in
disordered systems.  XAFS is an important technique for very short-range structure.\cite{prinz;b;xafs88}  Here we describe insights that we have gained from using the
atomic pair distribution function (PDF) analysis of powder neutron and x-ray diffraction 
data.\cite{egami;b;utbp02}

This technique is described in detail elsewhere.\cite{egami;b;utbp02}  Here we mention only that it yields
quantitative information about atomic structure on short (nearest neighbor) and intermediate (up to $\sim 10$~nm) length-scales.  The PDF is obtained by a sine Fourier transform of properly corrected and normalized x-ray or neutron powder diffraction data:\cite{egami;b;utbp02}
\begin{equation}
G(r) = 4\pi r[\rho (r) - \rho_0] = {2\over\pi}\int_{0}^\infty 
Q[S(Q) - 1]\,\sin Qr \, dQ,
\label{eq;pdf}\end{equation}
where $\rho (r)$ is the microscopic pair density, $\rho_0$ is the 
average number density, $S(Q)$ is total structure function that is
the normalized scattering intensity, and $Q$ is the magnitude of the
scattering vector, $Q = |{\bf k}-{\bf k_0}|$.  For elastic scattering,
$Q = 4\pi\sin\theta/\lambda$, where $2\theta$ is the scattering angle
and $\lambda$ is the wavelength of the scattering radiation.

New insights have been gained into the complex oxides by applying this approach.  In
particular, here we give an overview of results from cuprate superconductors and
colossal magnetoresistant manganites.  The role (or otherwise) of the structure in the properties
of these materials has been hotly debated over the years.  The strong electron-electron 
interactions and resulting magnetism clearly are important; however, different views are
held about the fundamental importance or otherwise of the lattice to the superconductivity
and magnetoresistance of these materials.  Although phonons were dismissed early as the sole
 mediators of pairing in the cuprates, a number of studies suggested subtle lattice effects 
 occurring.\cite{egami;pms94,egami;b;pphtsv96} though contradictions and disagreements between the results from different techniques made these hard to assess and interpret. In the manganites, because of the active Jahn-Teller distortion, the magnitudes of the structural effects
are much larger and there is much better agreement between the techniques in this case.  In
fact, lessons learned from manganite research is resulting in renewed efforts and better 
understanding of the situation in the much more subtle cuprates.  For this reason, we begin with the manganites.  This is not an extensive review of the field of lattice effects in the cuprates, but an overview of the discoveries that our group has made using the PDF technique.

\section{Manganites}\label{sec:mang}
\subsection{Polarons and colossal magnetoresistance}
As early as the 1950's the general relationship between magnetism and charge transport was
elucidated through Zener's double-exchange mechanism.\cite{zener;pr51}  The important role of
the Jahn-Teller effect was noted early on by Goodenough,\cite{goode;pr55} but its importance to the CMR phenomenon
was only properly elucidated in the mid 1990's by Millis~\etal ,\cite{milli;prl95} and Bishop and co-workers.\cite{roder;prl96}  The PDF provided some of the best experimental evidence that
charges were localizing as lattice (and spin, but the PDF only sees the lattice) polarons at
the metal-insulator transition, \tmi .\cite{billi;prl96} The evidence appeared in the PDF as an
anomalous broadening of the PDF peaks associated with \tmi .  The nature of the polaronic state 
was also studied at this time by modelling the structural changes evident in the PDF at \tmi 
.\cite{billi;prl96,billi;prb00}  The positive charge carriers (holes) localize on Mn\sp{4+} sites with a shrinking of the 
MnO\sb{6} octahedron to a small, regular octahedral shape.  The octahedra without a localized 
charge take on a Jahn-Teller elongated shape with a particularly long bond of $r=2.15$~\AA , compared with $r=1.91 - 1.96$~\AA\ for the short Mn-O bonds present.\cite{billi;prb00}  The appearance of polarons in the insulating phase is unambiguous in the local structure measurements.\cite{billi;prl96,billi;prb00,booth;prb96,louca;prb97,louca;prb99}
\subsection{FM phase is nanophase-segregated}
\begin{figure}[th]
\centerline{\epsfxsize=3.5in\epsfbox{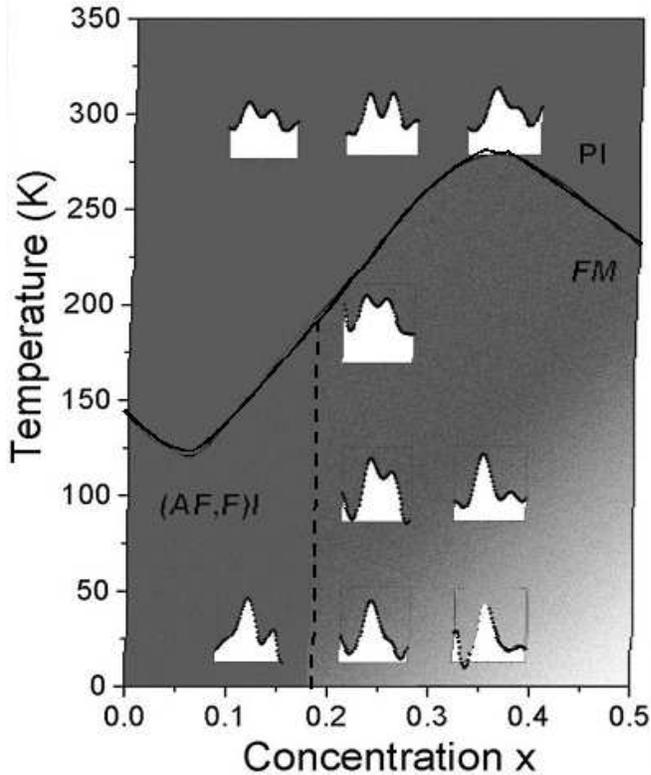}}   
\caption{Phase diagram of \protect\lcmo .  Superimposed are PDFs of the low-$r$ region measured using high-energy x-rays.  The peak at lower-$r$ is at the position of the short Mn-O bonds in an MnO\sb{6} octahedron, the higher-$r$ peaks in the doublets are at the position of the Mn-O long-bonds in JT distorted octahedra.  Clearly JT distorted octahedra are present throughout
the FM region of the phase diagram, except at high doping ($x>0.25$) and low-temperature. They grow in number as \tmi\ is approached. \label{fig;phased}}
\end{figure}

The next phase was to understand how these objects appeared.  Do the octahedra uniformly grow
with increasing temperature as \tmi\ is approached or do fully distorted polarons locally appear 
and grow in number?  A high resolution x-ray PDF measurement,\cite{billi;prb00} strongly suggested the latter.  In agreement with supporting neutron measurements it seemed that long bonds always appear at the distance expected for unstrained JT distorted octahedra (as seen in undoped LaMnO\sb{3} for example) and the number of such states increases with temperature towards \tmi .  This is shown in Fig.~\ref{fig;phased}.  

Interestingly, a significant number of JT distorted octahedra are evident well below \tmi\ in the ferromagnetic metallic (FM) phase.  Localized electronic states are coexisting with delocalized metallic states in this region.  This
suggests a phase separation into conducting and insulation regions and an inhomogeneous electronic structure.  Similar behavior was directly observed from dark field imaging TEM studies of a related manganite.\cite{uehar;n99}  However, in that case the insulating and metallic phases were micron sized and static.  In this case of the high-\tc\ \lcmo\ system {\it there is no diffraction or TEM evidence of macroscopic phase separation}.  Nonetheless, we are forced to the
conclusion from the PDF results that the samples are phase separated on a nanometer length-scale,
either statically or dynamically, and that this nanostructure evolves with temperature as the phase transition is approached.  This general picture in this, and related, manganites is now supported by range of experimental
evidences.\cite{moreo;s99}
\subsection{Percolation transition}
\begin{figure}[th]
\centerline{\epsfxsize=3.0in\epsfbox{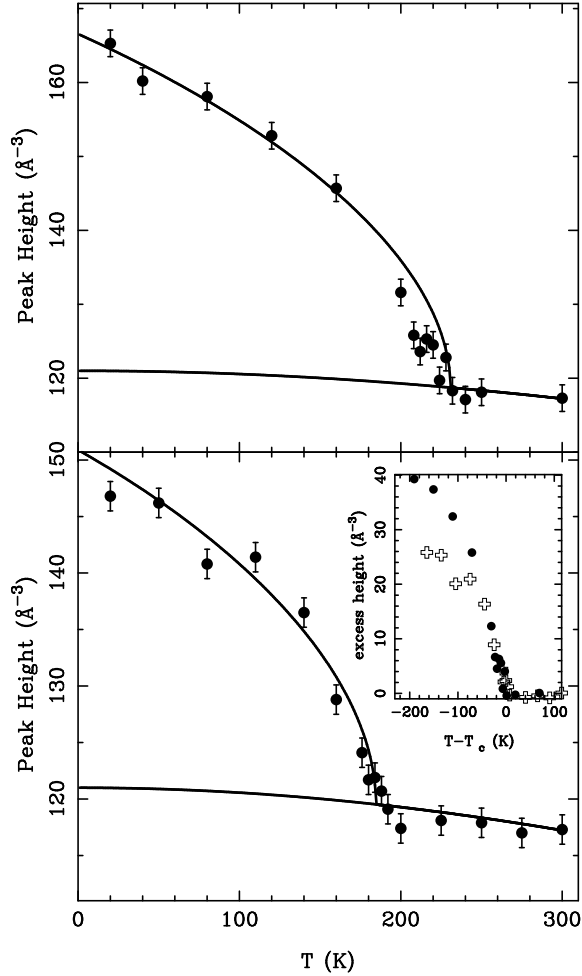}}   
\caption{PDF peak height vs. temperature for the peak in \lcmo\ at $r=0.275$ that originates from O-O
correlations on the MnO\sb{6} octahedra. The panels show the $x=0.25$ (top) and $x=0.21$ (bottom) compositions.  The lower solid line is the expected Debye behavior fit to the high-T region of the curve, the upper solid line is a guide to the eye.  In the inset is shown the
excess peak-height (data-points minus the Debye curve in each case) for both the $x=0.25$ and $x=0.21$ samples plotted vs. reduced temperature.  They exhibit scaling behavior, though the 
$x=0.21$ sample does not become fully delocalized even at low temperature so does not follow the $x=0.l25$ curve all the way.\label{fig;scaling}}
\end{figure}

\begin{figure}[th]
\centerline{\epsfig{file=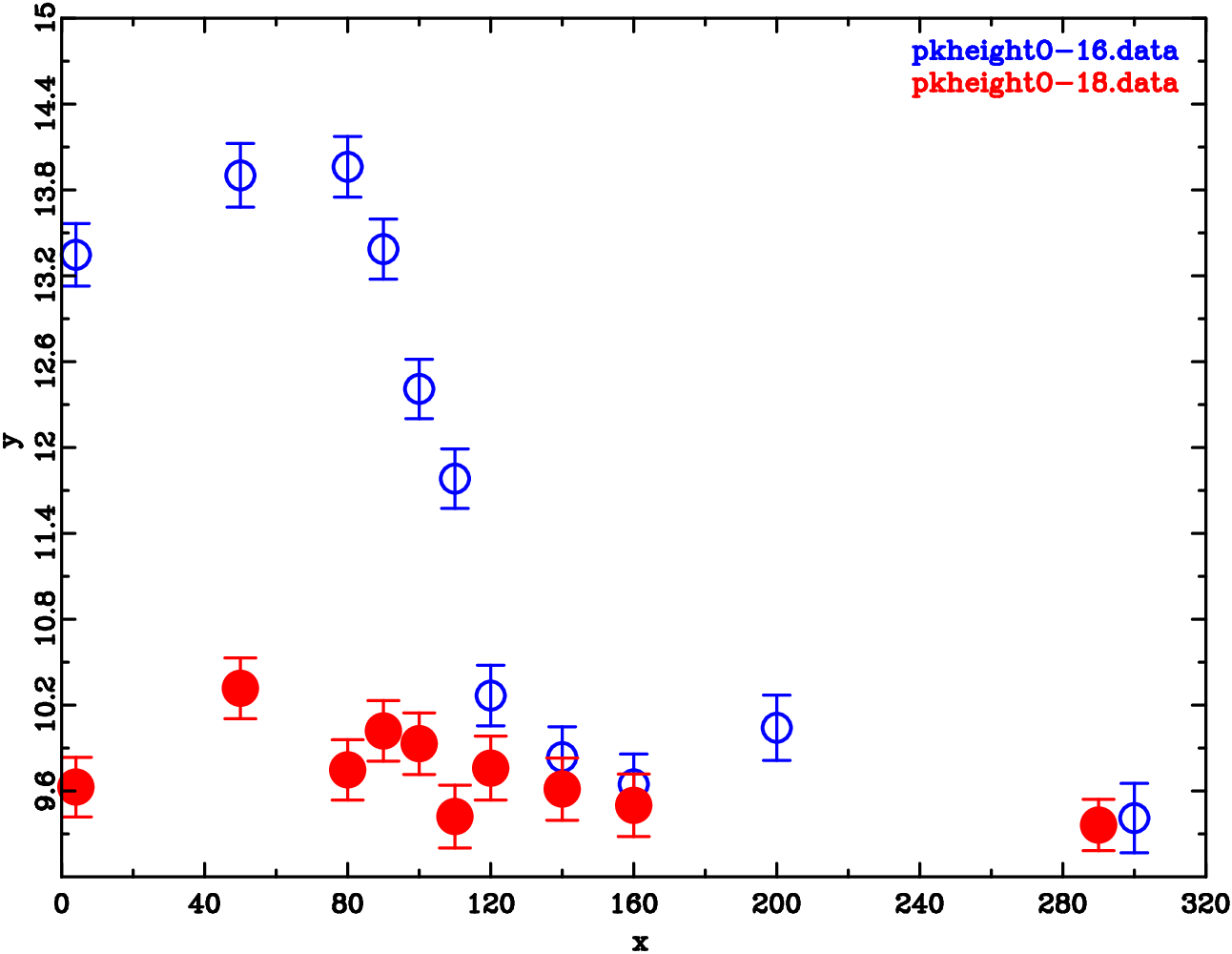,width=6.0cm,clip=TRUE,angle=-90}}
\vspace{0.2cm}   
\caption{Plot of PDF peak-height  of the $r=0.275$ O-O peak (arbitrary units) vs. temperature for two samples of La\sb{0.525}Pr\sb{0.175}Ca\sb{0.3}MnO\sb{3} enriched with $^{16}$O (open circles) and $^{18}$O (filled circles) respectively. In the $^{18}$O the transition to a FM ground-state is suppressed and the sample remains an AF insulator to low temperature.  This clearly shows that the excess PDF peak-height is coming from the appearance of undistorted octahedra in the metallic phase.\label{fig;oisotope}}
\end{figure}

This picture suggests that the MI transition may occur by a percolation mechanism so this was
investigated.  When the JT distorted octahedra appear the PDF peak at 2.75~\AA , coming from the
O-O bonds on the MnO\sb{6}octahedron, broadens.\cite{billi;prl96}  Since the PDF is a bulk
measurement, this gives us a semi-quantitative measure of how many octahedra are in the metallic
phase.  The peak width is inversely proportional to the peak height, hence 
$h \propto\left( {1\over\langle u^2\rangle}\right)^{0.5}$.  
As the sample goes into the metallic phase
polarons disappear, $\langle u^2\rangle$ decreases and the peak-height increases.  We were able to show that this behavior scales with reduced temperature, as shown in Fig.~\ref{fig;scaling}. 
These changes in peak height are really coming from the loss of polaronic behavior.  In the T-dependent studies we had to subtract the normal T-dependence (the Debye-Waller behavior) before
revealing the peak-height scaling shown in Fig.~\ref{fig;scaling}.  Recently, we have measured
a particular isotopically substituted ($^{18}$O for $^{16}$O) sample where the transition into
the FM phase is suppressed by the isotope substitution.  The results are shown in Fig.~\ref{fig;oisotope}.  The abrupt increase in PDF peak-height on entering the metallic phase is unambiguous.  

It is dangerous to call this an order parameter since this tends to imply second order behavior
that is not evident here.  However, it is a direct experimental measure of the degree to which
a sample is in the metallic phase.  It can be compared to mixing parameters derived from transport measurements\cite{jaime;prb99} and used in percolative models for transport.\cite{mayr;prl01}  Despite it not being an order parameter, Fig.~\ref{fig;scaling}
suggests that it shows universal-like behavior.  This is also supported by analysis of data from an \lcmo\ $x=0.5$ sample described below.  This behavior seems interesting and begs a theoretical explanation.
Note also that no macroscopic phase separation has been observed in the \lcmo\ system described here and by Jaime \etal\cite{jaime;prb99} and Mayr~\etal.\cite{mayr;prl01}  This places the current system somewhere between conventional notions of first and second order behaviors with apparently both kinds of behavior being observed depending on the length-scale and time-scale of the measurement; a classic multi-scale problem reminiscent of discussions about the phase transitions of BaTiO\sb{3}.\cite{kwei;jpc93}  Actually, in the \lcmo\ system a crossover from
first order to second order behavior has been postulated from specific heat measurements as a function of doping, with first
order behavior noted below $x<0.4$.\cite{kim;cm02}

\begin{figure}[t]
\centerline{\epsfig{file=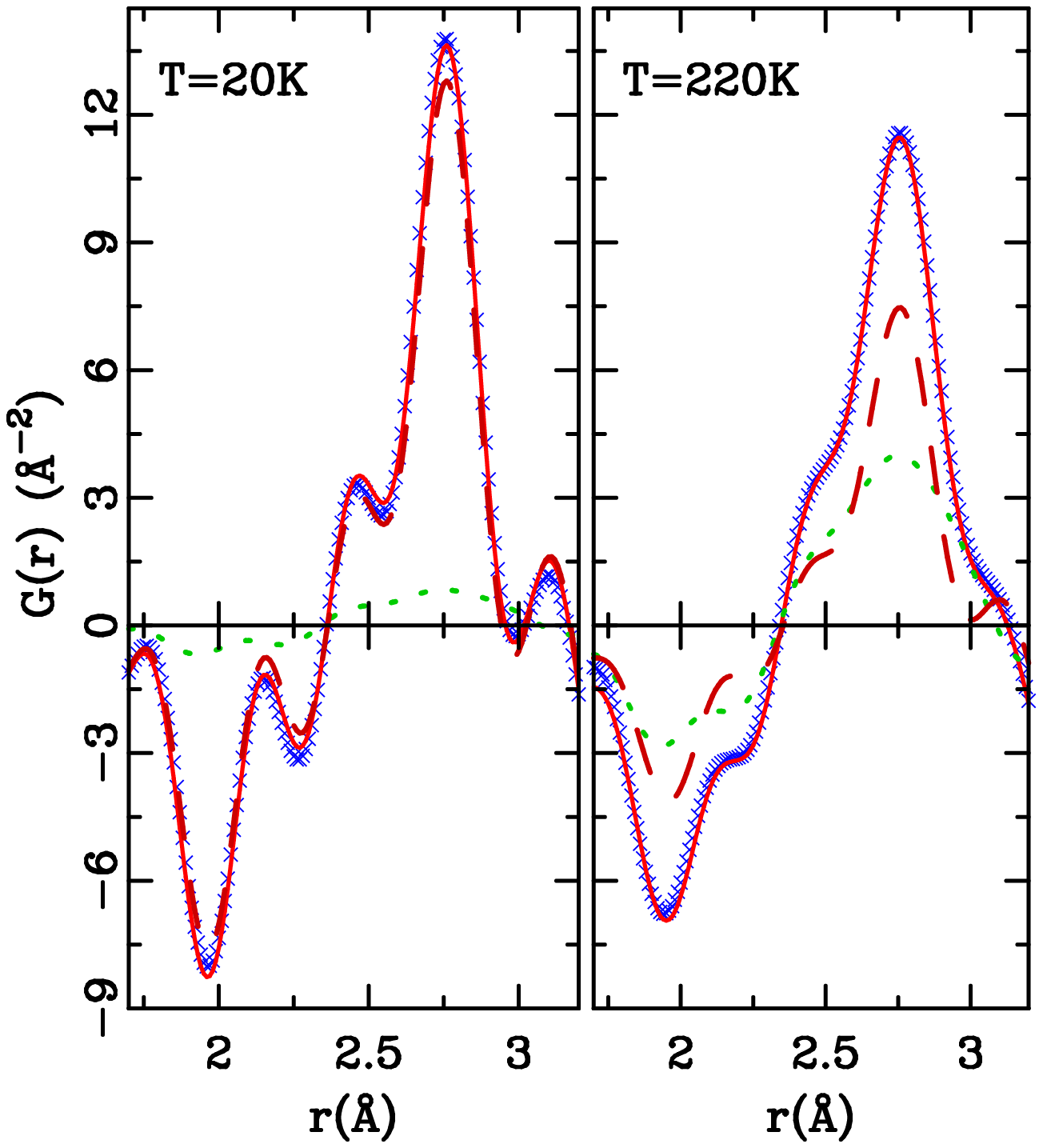,width=5.5cm,clip=TRUE,angle=0}
\epsfig{file=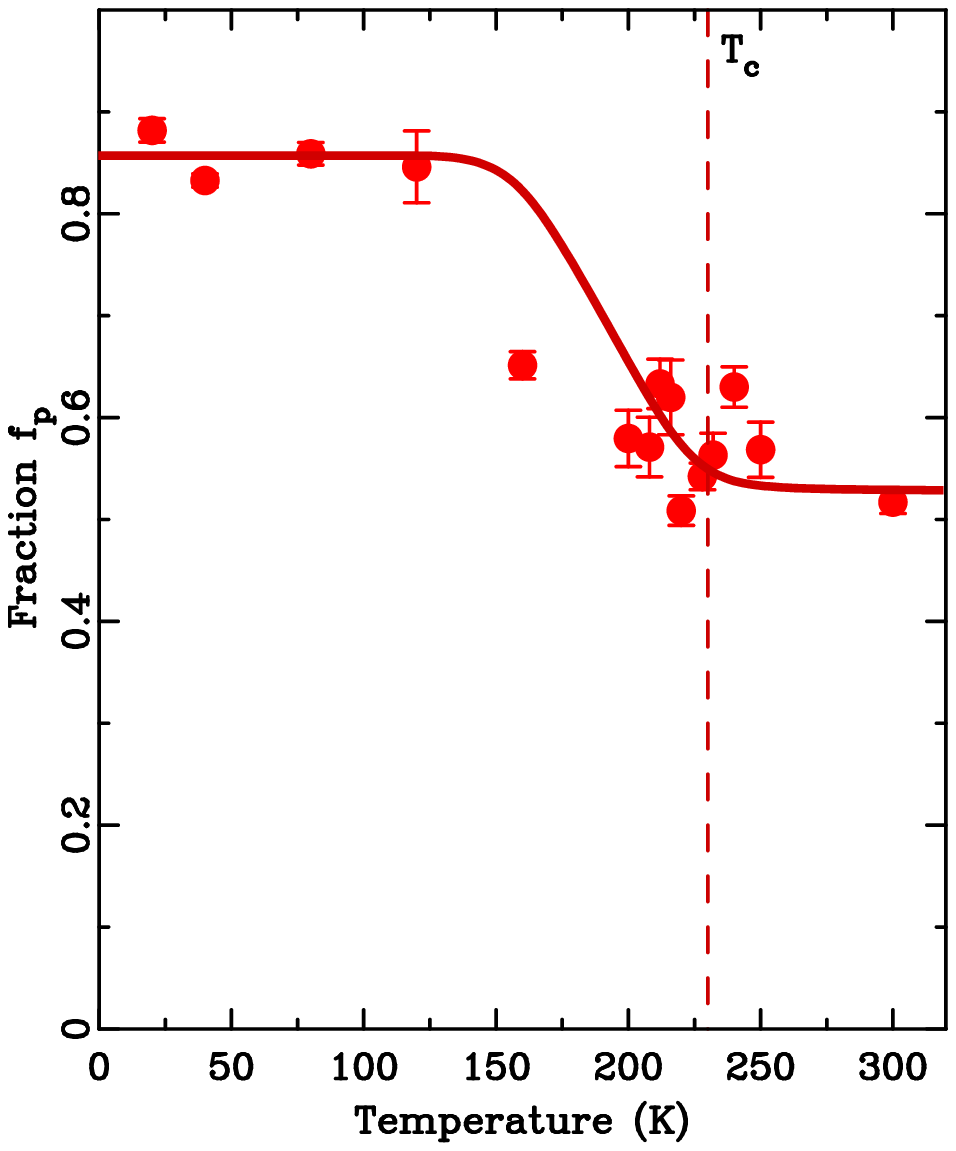,width=5.0cm,clip=TRUE,angle=0}
}
\vspace{0.2cm}   
\caption{First two panels: low-r region of \protect\lcmo\ $x=0.25$. Shown are data (symbols)
best-fit of two-phase models (solid line) distorted phase (dotted line) and undistorted
phase (dashed line). At low temperature (20K) the sample is predominantly undistorted, at
220K there is a significant proportion of both distorted and undistorted. Third panel:
the resulting refined phase fractions as a function of temperature. \label{fig;2phasefit}}
\end{figure}

The problem with the PDF peak-height parameter is that it doesn't yield the absolute value of the
metallic fraction but rather only tracks the {\it change} of this parameter with temperature (or doping, for example).  We have attempted to extract quantitatively the absolute
fraction of undistorted octahedra
by modelling the PDF.  Models for the undistorted and distorted phases are taken
from the \lcmo\ $x=0.3$ and $x=0.0$ samples, respectively, at low temperature.  We are
interested in the proportion of distorted and undistorted octahedra rather than how they
arrange in space so the PDF is fit only over a range to $\sim 4$~\AA .  A characteristic fit is 
shown in Figs.~\ref{fig;2phasefit}(a) and (b) and the resulting phase fractions are shown in
Fig.~\ref{fig;2phasefit}(c).  These indicate that this sample ($x=0.25$) is more than 95\%
in the metallic phase at low temperature but, surprisingly, remains at least 50\% undistorted
above \tmi .  This appears
to be supported by neutron small angle scattering measurements which see significant nanometer-scale ferromagnetic clusters surviving far above \tmi .\cite{radae;unpub01}  
A word of caution is necessary here.  The 
phase fractions determined from the PDF fitting will depend on the structural models used
for the two phases.  A more appropriate model for the distorted polaronic phase may be the
polaronic charge ordered phase observed at low-temperature in the $x=0.5$ sample.\cite{radae;prb97}  These results are interesting in light of a possible percolation model for the transport because, (a) If the sample is 50\% metallic above \tmi , this already exceeds the
geometric percolation threshold for the 3-D cubic lattice (though not the 2-D case) suggesting
non-random percolation and (b)
the sharp increase in the number of undistorted octahedra (associated with the metallic state) at
\tmi\ would suggest a strong feedback mechanism. The metallic phase increases slowly with decreasing temperature until it percolates whence it begins to increase rapidly (e.g., see 
Fig.~\ref{fig;scaling}). Again, this suggests some kind of correlated percolation if the 
percolation picture is right at all.  The origin of this behavior is not completely clear,
but clearly the electronic properties depend on structures on multiple scales.

\begin{figure}[t]
\centerline{\epsfig{file=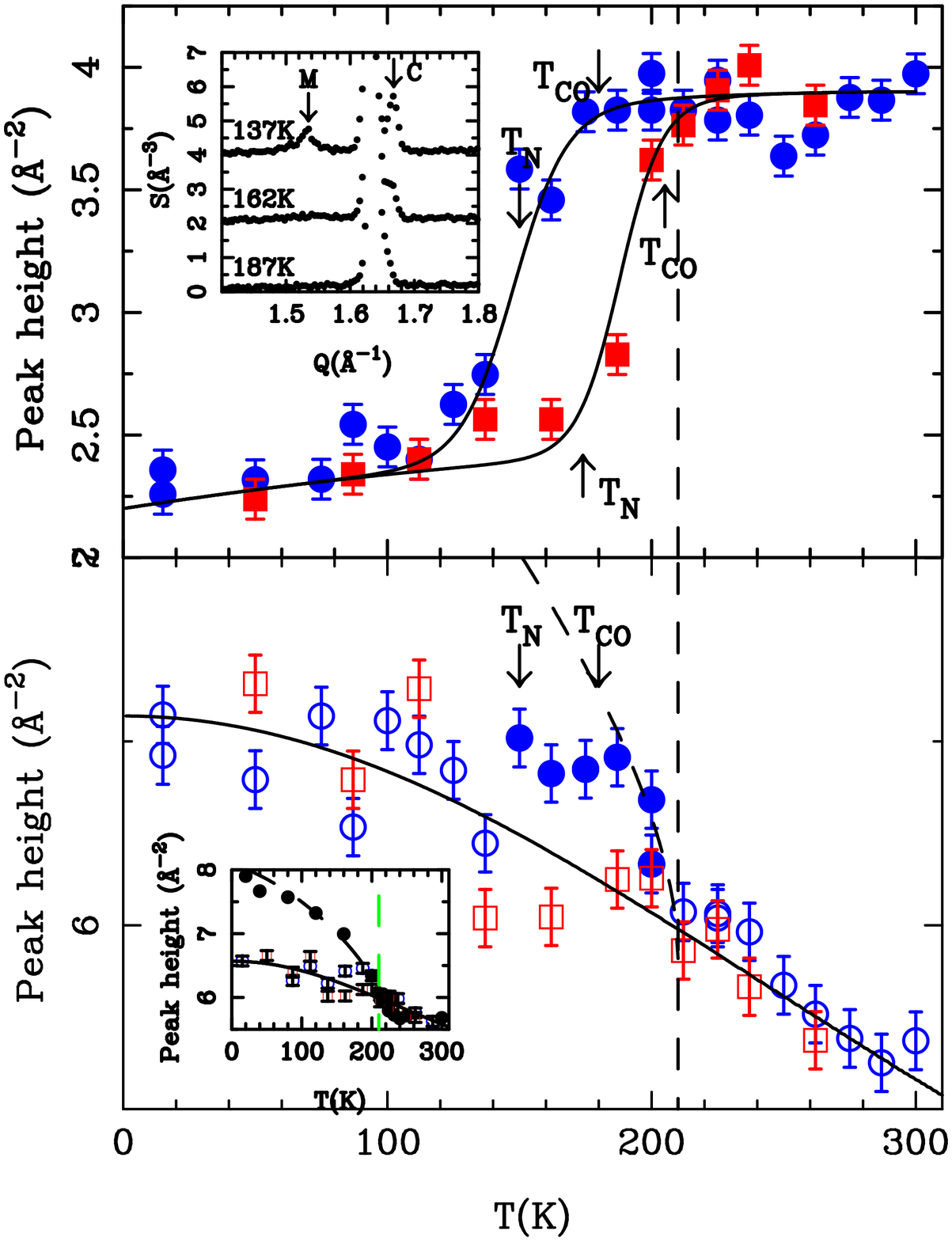,width=6.0cm,clip=TRUE,angle=0}}
\vspace{0.2cm}   
\caption{50\% PDF peakheight plots from \lcmo\ $x=0.5$ as a function of temperature.
The top panel shows a high-$r$ peak that is sensitive to the charge-ordering
transition (indicated by arrows for cooling and warming) but not the ferromagnetic transition (shown by the dashed line).  The lower curve shows the O-O peak at $r=0.275$~\AA\ that responds to the FM transition by sharpening, following the scaling curve discussed above and shown as a dashed line.  However, when the portion of the sample that is still polaronic charge orders,
the delocalized portion of the sample is quickly swallowed up the peak-height returns to the
smooth curve of the distorted phase. This is a failed MI transition. \label{fig;0.5}}
\end{figure}

A similar competition between localized, polaronic and delocalized metallic behavior, 
is occurring in the $x=0.5$ sample.  Yet another length-scale comes into play in this case.
In heavily doped materials the polarons prefer to order to minimize strain; so called 
charge order. In \lcmo\ at $x=0.5$ this occurs in zigzag stripes that results in CE 
magnetic order\cite{goode;pr55} so we will refer to it as CE charge order.  The ground-state
at $x=0.5$ is  antiferromagnetic insulating with CE charge order.  However, this sample is
very close to the FM phase at slightly lower doping and, indeed, on cooling the sample
goes first ferromagnetic before becoming charge ordered at lower temperature.  We have
plotted the PDF peak height through these transitions as shown in Fig.~\ref{fig;0.5}.\cite{billi;prl99;unpub}
On becoming ferromagnetic the $r=2.75$~\AA\ peak that showed the scaling behavior
(Fig.~\ref{fig;scaling}) again sharpens on going through the ferromagnetic transition,
following the scaling law which is plotted as a dashed line in Fig.~\ref{fig;0.5}; the
sample is heading towards its ferromagnetic ground-state and gradually transforming
into undistorted, delocalized metal.  Nonetheless, much of the sample still remains polaronic.
At $T=180$~K this polaronic part charge-orders, apparent from the appearance of CO superlattice
peaks at this temperature.  Below this temperature, instead of the PDF peak-height curve
following the scaling law curve towards complete metallicity it drops back to the average
line (Fig.~\ref{fig;0.5}) and the part of the sample already transformed to metallic is restored
to a polaronic state.  The energy balance between these two states is very delicate apparently
and the lowering in energy of the polaronic state due to it being permitted to order commensurately with the lattice is enough to stabilize it with respect to the FM phase.
The length-scale for the charge order is roughly the stripe-separation which is around 10~\AA .
This is larger than the $\sim 3$~\AA\ size of the polarons themselves and the nanometer
length-scale (imprecisely determined) of the nano-phase separation.  This picture is also
consistent with the observation of short-range ordered charge-ordered clusters in the polaronic
phase above \tmi\ at dopings below $x=0.5$.\cite{lynn;jap01}

\section{Cuprates}\label{sec:cupr}

\subsection{Stripes and structural distortions}
Our investigations in the cuprates have focussed principally on the \lsco\ system.
Here the PDF has been used to look for evidence of localized charges in direct analogy
with the case of the manganites.  The polaronic distortions in this case are thought to be
much smaller, being somewhere between $\sim 0.02$~\AA\cite{bozin;prl00} and $\sim 0.004$~\AA\cite{tranq;prb96} making them 10-50 times smaller than in the manganites (though
note that XAFS data have been interpreted as due to a larger distortion on a minority of
sites.\cite{bianc;prl96})  The small size of polaronic distortions in this system makes them difficult to
study using structural probes and results should be considered only semi-quantitative.
The size of the polaronic distortion predicted by Bozin~\etal\cite{bozin;prl00} is based
not directly on PDF measurements but taken from the magnitude of bond-shortening observed
in the average structure due to doping and determined crystallographically.  This is beyond
the resolution of the PDF (and XAFS) which is around $r=0.12$~\AA\ and set by physical limits;
i.e., the quantum zero-point motion of the atoms.  Differences in bond-length shorter than this
cannot be seen directly in PDF and XAFS data but only inferred by multi-peak fits to a single
feature.  

Intermediate range data is also present in powder diffraction derived PDFs.  This 
contains information about such things as CuO\sb{6} octahedral tilts. The presence of polarons
or stripes has implications on the octahedral tilts and this can give complementary information
about the presence or absence of such disorder.  Below we briefly describe the PDF evidence
for local short-range ordered stripes, though this has been extensively described in a series of
publications.\cite{bozin;prl00,bozin;prb99}

\subsection{Structural evidence for short-range ordered stripes}

\begin{figure}[t]
\centerline{\epsfig{file=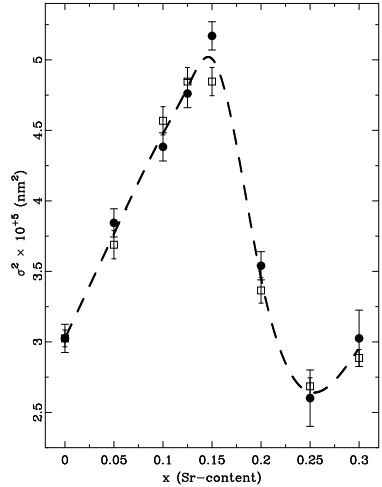,width=5.0cm,clip=TRUE,angle=0}}
\vspace{0.2cm}   
\caption{Peak width of the in-plane Cu-O PDF peak as a function of doping at 10K.
\label{fig;pkbroad}}
\end{figure}

First we develop an argument about the structural consequences of charge stripes, then we
look in the PDF data for supporting evidence.  The first observation is that doping charge
into the planar Cu-O shortens them.  This is a universally observed experimental fact in the
cuprates and is easily understood since the planar bonds are $\sigma^*$ antibonding bands and
doping positive charge (holes) therein stabilizes the covalent bond and shortens it.  Based
on data of Radaelli~\etal\cite{radae;prb94i} the (average) bond shortening in \lsco\ is 
${d r_{Cu}\over d x}\sim -0.1$~\AA/doped charge/copper, thus doping $x=0.2$ leads to
a bond shortening of $\sim 0.02$~\AA .  Now we assume the presence of stripes.  This means
1-D stripey objects with increased doped charge on them separated by stripes with less doped
charge.  It immediately becomes obvious that the ``charged stripes" must have shorter bonds than the
``uncharged stripes" though they are topologically connected by covalent bonds.  This
has experimental and theoretical implications.  The theoretical implications are that the existence of charge stripes introduces a misfit strain that tends to break the stripes up.  This is discussed below.  The experimental implications are that evidence should exist in structural
probes for a distribution of planar Cu-O bond lengths.  When the stripes are long-range ordered
this will appear in the form of superlattice peaks in neutron diffraction as already observed.\cite{tranq;prb96}  If the stripes exist locally but are not long-range ordered we
have to look using a probe sensitive to the local structure such as PDF.  As
already mentioned this difference in bond length cannot be directly observed so it is necessary
to search for a doping dependent broadening in the planar Cu-O bond length distribution.  This
was observed in \lsco ,\cite{bozin;prl00} peaking at $x=0.15$ (rather than at $x=0.5$ which would
be expected if the charges are localized randomly on single copper sites).  This is shown in Fig.~\ref{fig;pkbroad}(a).
Above $x=0.15$ the
bond distribution sharpened again, presumably as the stripe order is replaced by a more homogeneous charge distribution.  Temperature dependence of the bond distribution has also been
studied.  There is much scatter in the data, which testifies to the difficulty of the measurements
due to the small polaronic distortion, but an anomalous broadening at low temperature (below $\sim 100$~K) is evident\cite{gutma;unpub00} that suggests the appearance of charge stripes.  This is consistent with XAFS data that shows
a broadening in the planar Cu-O bond distribution at low temperature,\cite{bianc;prl96} although
the modelling of these data resulted in a different interpretation.\cite{bianc;prl96}

There are other structural implications of the presence of stripes.  The CuO\sb{2} planes
in \lsco\ are buckled by cooperative tilts of CuO\sb{6} octahedra.  You can shorten a Cu-O bond
locally without straining it longitudinally by locally removing the octahedral tilt.  The existence of short-bond charged stripes therefore also implies that there will be a distribution
of CuO\sb{6} octahedral tilt {\it angles}.  The intermediate range region of the measured PDFs should be 
consistent with this if stripes are present.  Indeed, a model that contained mixed tilt amplitudes reproduced the data very well\cite{bozin;prb99,bozin;prl00} which, at least, shows that the observed PDFs are consistent with the presence of octahedral tilt disorder.  Furthermore, this study\cite{bozin;prb99,bozin;prl00} indicated that some tilt {\it directional} disorder was also
present (i.e.,  a mixture of $\langle 110 \rangle$ and $\langle 100 \rangle$ symmetry tilts).
This is expected from topological arguments.\cite{billi;ijmpb02}  The PDF data are therefore
consistent with the presence of short-range ordered charge-stripes in the underdoped
region of \lsco\ at low temperature.

\subsection{Multi-scale structure: stripe domains}
 
We new briefly describe the theoretical implications of the above discussion, and that is the
appearance of an ``interfacial" lattice misfit strain as a  direct consequence of charge-stripe
formation.  The basic arguments that lead to this conclusion were laid out at the beginning of 
the last section and have been described in detail elsewhere.\cite{billi;prb02,billi;ijmpb02}
When charge is doped into the CuO\sb{2} plane the bonds shorten.  The presence of stripes implies
that regions of the plane coexist side-by-side that are heavily doped and lightly doped.
These regions are topologically connected by covalent bonds.  The charged stripes want to be
shorter than the uncharged regions between and this is the origin of the misfit strain.  As
the stripe gets longer the strain increases.  An infinite stripe would have an infinite strain
energy, so at some characteristic length that is a balance between the stripe formation
tendency (presumed to come from the electronic system and magnetism) and the strain energy and
the stripes break up.
A simple model that captures this physics is described in Refs.~\cite{billi;prb02,billi;ijmpb02}.  A lattice gas with attractive near neighbor ($J_{nn}$)
and repulsive next nearest neighbor ($J_{nnn}$) interactions for doped sites ensures stripe formation.
The strain terms come about due to a misfit of the desired bond-length of the doped site and the
constraint imposed by the average periodic potential that it sits in, determined by the rest of the crystal.

The critical breakup length for the stripes is given by $L_c=N_ca$, where
the critical number of copper sites, $N_c$, per
strained stripe scales like
\begin{equation}
N_c\sim\left( {J_{nn}\over k_{nn}} \right) \left( {1\over a-l_0 } \right)^2 .
\label{eq;nc}
\end{equation}
Here, $k_{nn}$ is the harmonic spring constant between nearest neighbor doped sites, $a$ is the
average separation of copper ions dictated by the average structure and $l_0$ is the
bond length of the shorter, doped, Cu-O bonds in the stripe.  

\begin{figure}[t]
\centerline{\epsfig{file=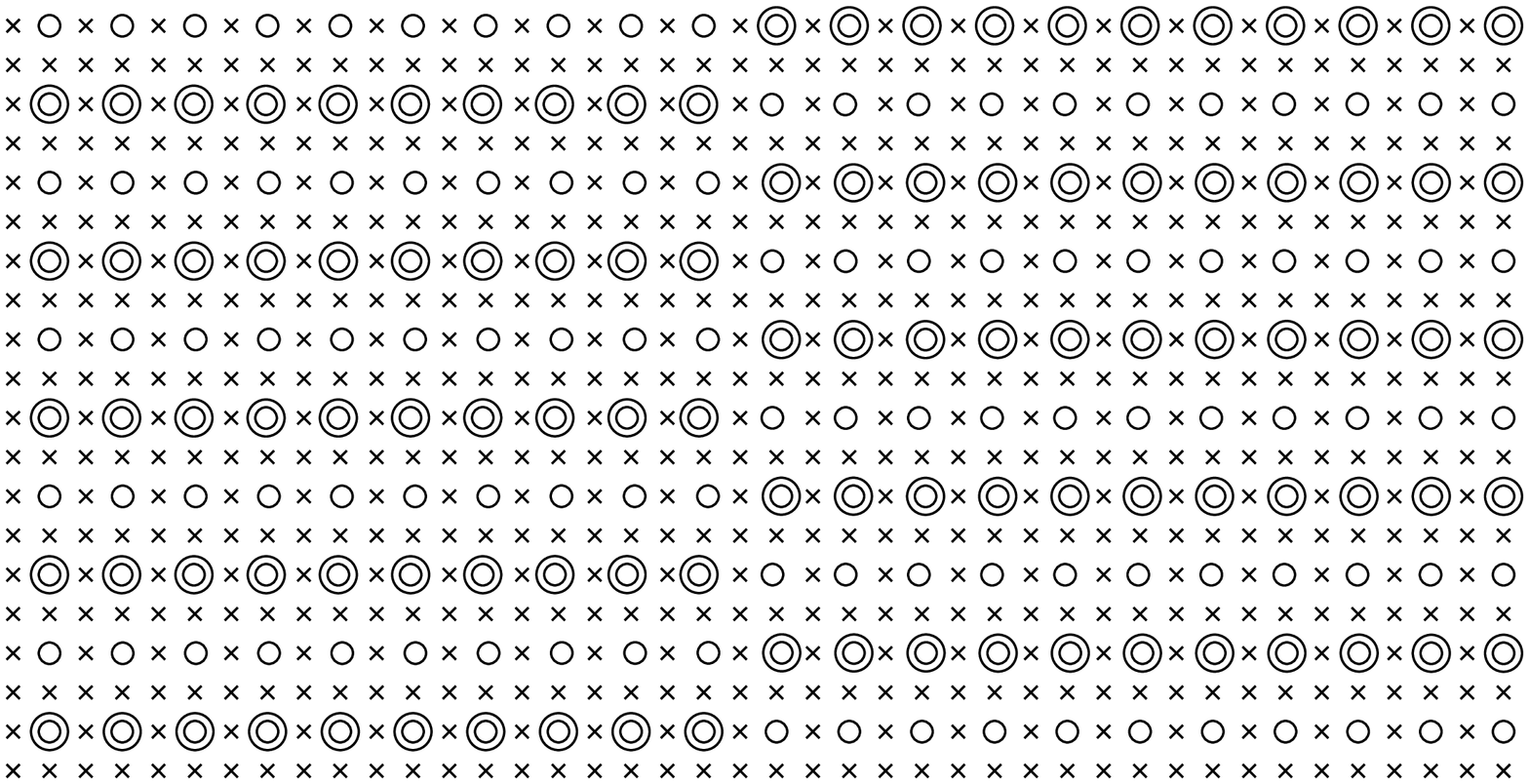,width=5.5cm,clip=TRUE,angle=-90}
\epsfig{file=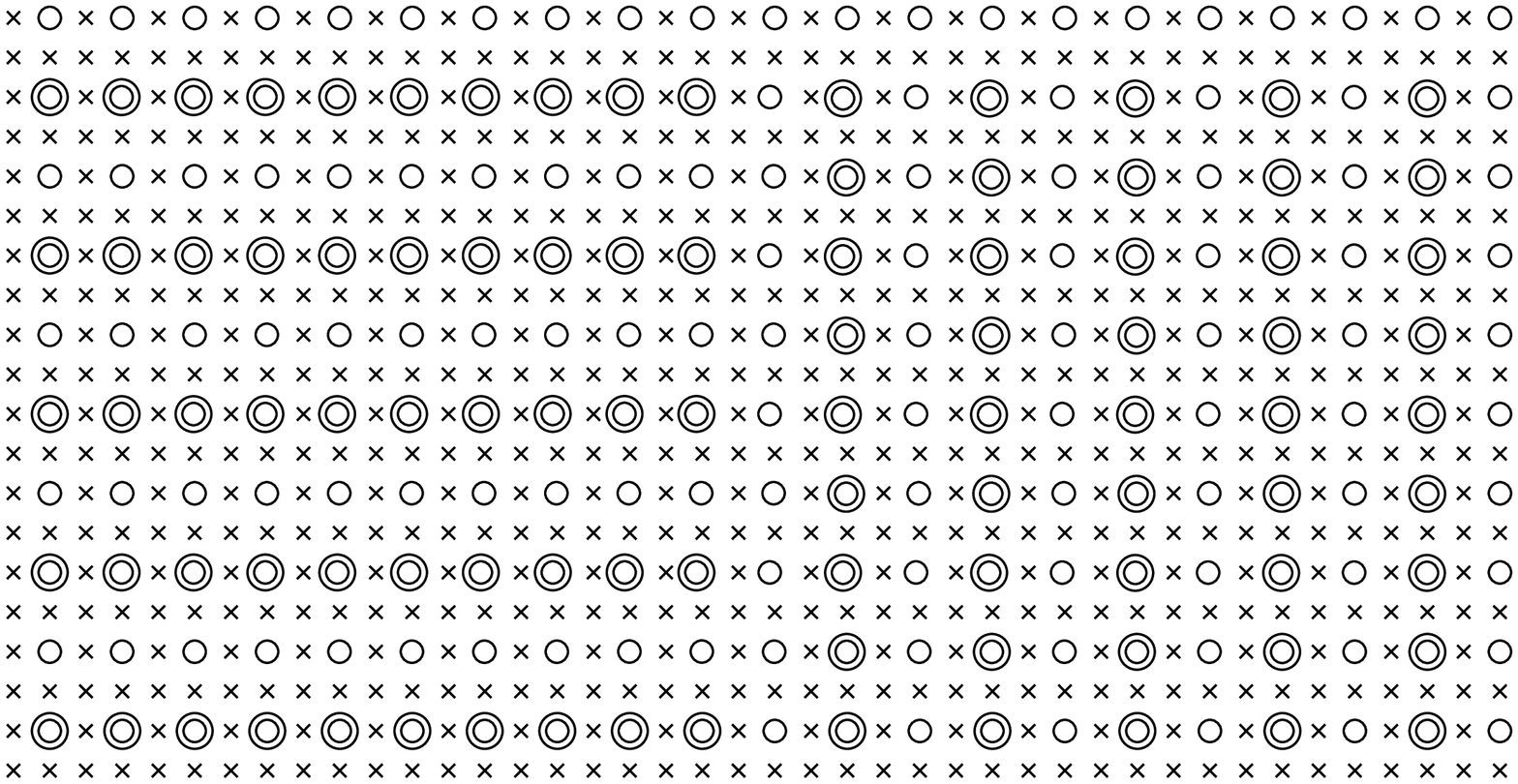,width=5.5cm,clip=TRUE,angle=-90}}
\vspace{0.2cm}   
%
\caption{Stripe nanostructures in cuprates.  (a)  Interleaved
stripes (b) weave microstructure. One unit
cell of each microstructure is shown. Concentric circles indicate doped
sites, while sites indicated by crosses are excluded from the strain relaxation.
 For illustration purposes the natural length of bonds  between doped sites are
10\% shorter
than bonds between undoped sites. }
\label{fig;stripemicros}
\end{figure}

This inherent tendency towards stripe breakup due to lattice strain will give rise to 
microstructures with domains of broken stripes.  A number of intuitive possibilities are shown
in Fig.~\ref{fig;stripemicros}.
%
%
%
The first relieves strain by matching up short charged stripes with longer uncharged stripes.
This has significantly lower energy than the system without the broken stripes.  It has
similar energy to that shown in Fig.~\ref{fig;stripemicros}(a) that has the additional advantage
that it is more isotropic and allows better strain relaxation in two dimensions.  It should
be noted that details of the magnetism, not included in this model, will impact the preferred
microstructure because of spin frustration in the interfacial regions between domains. However,
the general picture that strain breaks up stripes into a microstructure with a characteristic
length-scale that is given approximately by Eq.~\ref{eq;nc} is robust.

This general picture may explain some of the phenomenology of the cuprates.  If we assume that
static, long-range ordered, stripes compete with superconductivity but short-range ordered 
dynamic stripes do not, or that fluctuating stripes even enhance superconductivity,\cite{emery;pnas99} we can make the following observations.
A longer length-scale for stripe breakup will result in more slowly fluctuating stripes,
and a poorer superconductor, than a material with a shorter stripe breakup length-scale.
A shorter length-scale appears in systems with more strain.  At this point we haven't discussed
how this interfacial strain can be relaxed; however, a number of possibilities exist.
For example, in the \lsco\ system the CuO\sb{2} planes are buckled.  Part of the bond-shortening
required when a copper is doped can be accommodated by locally driving away the tilt without
straining the Cu-O covalent bond.\cite{billi;prb02}  Thus, we might expect that in this system the misfit strain that leads to stripe breakup is less than that in a system with flat planes, for example.  Longer stripes result, a larger microstructure and slower fluctuations.  In the 
extreme case where the sample is co-doped with misfitting Nd, or doped with misfitting Ba, there
may be sufficient structural compliance in the tilts to accommodate all the bond shortening allowing long-range ordered static stripes as observed in these systems, though not observed
in the Sr doped case.  Note that Sr\sp{2+} is a well-fitting replacement for La\sp{3+} and does
not perturb (i.e., increase) the octahedral tilt background too much.  The ability to accommodate
the bond shortening without a resulting misfit strain we call ``structural compliance".

There are other possible sources of structural compliance in the cuprates.  For example, in the
YBCO system, chain oxygen atoms have the opportunity to self organize so as to minimize the
misfit strain in the cuprate planes.  Chain-oxygen ordering has a well documented effect on T\sb{c} in these systems\cite{shake;prb95} though this was hitherto thought to be due to
charge-transfer effects.  Similar effects are seen due to interstitial oxygen ordering in
La$_2$CuO$_{4+\delta}$ which again may be related to self organization to minimize the energy
of the stripe microstructure.

A general observation is that as the CuO\sb{2} planes get flatter (and therefore the tilting source of structural compliance disappears), \tc\ goes up.  Focussing attention on the materials with a single CuO\sb{2} layer it is observed that the highest \tc\ material is HgBa\sb{2}CuO\sb{6}.  This has flat planes and a very simple structure with few possibilities for structural compliance.  Likewise, the single layer thallium compound which
is also a high-\tc\ material, has flat planes and just a bit of structural disorder in the
out-of-plane layers\cite{toby;prl90} that could self-organize.  Also a high-\tc\ material but
with a little lower optimal \tc , is the bismuth material that relaxes a mismatch between
the CuO\sb{2} and intergrowth layers with an incommensurate structural modulation.\cite{yamam;prb90}
YBCO is a two-layer system with a moderate \tc\ (similar to the single layer mercury compound
and the lowest of the two-layer bismuth, thallium or mercury materials) and it also has buckled
CuO\sb{2} planes.  

These empirical observations are at least qualitatively explainable within the picture
of  strain induced stripe-breakup and microstructure due to misfit strain.  They point to
the importance of engineering structures on multiple different length-scales in order to optimize
electronic properties in these materials.  Here it seems the atomic scale is (as always) important, but also the stripe length-scale of $\sim 10$~\AA\ and the length-scale of the
stripe microstructure that can vary from short (maybe comparable to the stripe spacing) all
the way to micron sized, with comparable change in material properties.

\section{Conclusions}
Here we have summarize investigations of the local atomic structure in the transition
metal cuprates and manganites.  Nano-scale electronic inhomogeneities appear to be widespread
and can have structures on a number of different length-scales.  These multi-scale structures
have a profound effect on the electronic properties of these materials.  In the case of
the cuprates we have presented a model for lattice-strain induced stripe breakup.  The lattice
strain is an inevitable consequence of having a microscopically inhomogeneous charge distribution
and has analogs in more systems which support variable doping and possible charge inhomogeneities.

\section*{Acknowledgments}
None of this work would have been possible without the tireless efforts of past and
present members of the Billinge group: Emil Bo\v{z}in, Matthias Gutmann, Thomas Proffen, 
Valeri Petkov, Peter Peterson, Il-Kyoung Jeong and Xiangyun Qiu.  It also benefitted
from financial support from NSF through grant DMR-0075149.  The results presented made use
of a number of x-ray and neutron facilities: IPNS at Argonne National Laboratory (DOE-BES contract number W-31-109-Eng-38), 
MLNSC, Los Alamos National Laboratory (DOE contract W-7405-ENG-36), APS, Argonne National Laboratory (DOE BES contract W-31-109-Eng-38).


\begin{thebibliography}{10}

\bibitem{salje;cp00}
E.~K.~H. Salje,
\newblock Contemp. Phys. {\bf 41}, 79 (2000).

\bibitem{sheno;prb99}
S.~R. Shenoy, T.~Lookman, A.~Saxena, and A.~R. Bishop,
\newblock Phys. Rev. B {\bf 60}, R12537 (1999).

\bibitem{stojk;prb00}
B.~P. Stojkovi\'c, Z.~G. Yu, A.~L. Chernyshev, A.~R. Bishop, A.~H.~C. Neto, and
  N.~Gr{\/ o}nbech-Jensen,
\newblock Phys. Rev. B {\bf 62}, 4353 (2000).

\bibitem{rasmu;prl01}
K.~{\/ O}. Rasmussen, T.~Lookman, A.~Saxena, A.~R. Bishop, R.~C. Albers, and
  S.~R. Shenoy,
\newblock Phys. Rev. Lett. {\bf 87}, 055704 (2001).

\bibitem{castr;prb01}
A.~H. {Castro Neto},
\newblock Phys. Rev. B {\bf 64}, 104509 (2001).

\bibitem{uehar;n99}
M.~Uehara, S.~Mori, C.~H. Chen, and {S.-W. Cheong},
\newblock Nature {\bf 399}, 560 (1999).

\bibitem{pan;n01}
S.~H. Pan, J.~P. O'Neal, R.~L. Badzey, C.~Chamon, H.~Ding, J.~R. Engelbrecht,
  Z.~Wang, H.~Eisaki, S.~Uchida, A.~K. Gupta, K.-W. Ng, E.~W. Hudson, K.~M.
  Lang, and J.~C. Davis,
\newblock Nature {\bf 413}, 282 (2001).

\bibitem{prinz;b;xafs88}
R.~Prinz and D.~Koningsberger, editors,
\newblock {\em X-ray absorption: principles, applications techniques of EXAFS,
  SEXAFS and XANES},
\newblock J. Wiley and Sons, New York, 1988.

\bibitem{egami;b;utbp02}
T.~Egami and S.~J.~L. Billinge,
\newblock {\em Underneath the Bragg Peaks: Structural analysis of complex
  materials},
\newblock Pergamon, Oxford, England, 2002.

\bibitem{egami;pms94}
T.~Egami and S.~J.~L. Billinge,
\newblock Prog. Mater. Sci. {\bf 38}, 359 (1994).

\bibitem{egami;b;pphtsv96}
T.~Egami and S.~J.~L. Billinge,
\newblock in {\em Physical properties of high-temperature superconductors V},
  edited by D.~M. Ginsberg, page 265, Singapore, 1996, World--Scientific.

\bibitem{zener;pr51}
C.~Zener,
\newblock Phys. Rev. {\bf 82}, 403 (1951).

\bibitem{goode;pr55}
J.~B. Goodenough,
\newblock Phys. Rev. {\bf 100}, 564 (1955).

\bibitem{milli;prl95}
A.~J. Millis, P.~B. Littlewood, and B.~I. Shraiman,
\newblock Phys. Rev. Lett. {\bf 74}, 5144 (1995).

\bibitem{roder;prl96}
H.~R\"oder, J.~Zang, and A.~R. Bishop,
\newblock Phys. Rev. Lett. {\bf 76}, 1356 (1996).

\bibitem{billi;prl96}
S.~J.~L. Billinge, R.~G. DiFrancesco, G.~H. Kwei, J.~J. Neumeier, and J.~D.
  Thompson,
\newblock Phys. Rev. Lett. {\bf 77}, 715 (1996).

\bibitem{billi;prb00}
S.~J.~L. Billinge, {Th.~Proffen}, V.~Petkov, J.~Sarrao, and S.~Kycia,
\newblock Phys. Rev. B {\bf 62}, 1203 (2000).

\bibitem{booth;prb96}
C.~H. Booth, F.~Bridges, G.~J. Snyder, and T.~H. Geballe,
\newblock Phys. Rev. B {\bf 54}, R15606 (1996).

\bibitem{louca;prb97}
D.~Louca, T.~Egami, E.~L. Brosha, H.~{R\"{o}der}, and A.~R. Bishop,
\newblock Phys. Rev. B {\bf 56}, R8475 (1997).

\bibitem{louca;prb99}
D.~Louca and T.~Egami,
\newblock Phys. Rev. B {\bf 59}, 6193 (1999).

\bibitem{moreo;s99}
A.~Moreo, A.~Yunoki, and E.~Dagotto,
\newblock Science {\bf 283}, 2034 (1999).

\bibitem{jaime;prb99}
M.~Jaime, P.~Lin, S.~H. Chun, M.~B. Salamon, P.~Dorsey, and M.~Rubinstein,
\newblock Phys. Rev. B {\bf 60}, 1028 (1999).

\bibitem{mayr;prl01}
M.~Mayr, A.~Moreo, J.~A. Verg\'{e}s, J.~Arispe, A.~Feiguin, and E.~Dagotto,
\newblock Phys. Rev. Lett. {\bf 86}, 135 (2001).

\bibitem{kwei;jpc93}
G.~H. Kwei, A.~C. Lawson, S.~J.~L. Billinge, and S.-W. Cheong,
\newblock J. Phys. Chem. {\bf 97}, 2368 (1993).

\bibitem{kim;cm02}
D.~Kim, B.~Revaz, B.~L. Zink, F.~Hellman, J.~J. Rhyne, and J.~F. Mitchell,
\newblock (2002),
\newblock cond-mat/0210088.

\bibitem{radae;unpub01}
P.~G. Radaelli and D.~Argyriou,
\newblock private communication.

\bibitem{radae;prb97}
P.~G. Radaelli, G.~Iannone, M.~Marezio, H.~Y. Hwang, S.-W. Cheong, J.~D.
  Jorgensen, and D.~N. Argyriou,
\newblock Phys. Rev. B {\bf 56}, 8265 (1997).

\bibitem{billi;prl99;unpub}
S.~J.~L. Billinge, R.~G. DiFrancesco, M.~F. Hundley, J.~D. Thompson, and G.~H.
  Kwei,
\newblock Phys. Rev. Lett.  (2000),
\newblock Unpublished.

\bibitem{lynn;jap01}
J.~W. Lynn, C.~P. Adams, Y.~M. Mukovskii, A.~A. Arsenov, and D.~A. Shulyatev,
\newblock J. Appl. Phys. {\bf 89}, 6846 (2001).

\bibitem{bozin;prl00}
E.~S. Bo{\v z}in, S.~J.~L. Billinge, H.~Takagi, and G.~H. Kwei,
\newblock Phys. Rev. Lett. {\bf 84}, 5856 (2000).

\bibitem{tranq;prb96}
J.~M. Tranquada, J.~D. Axe, N.~Ichikawa, Y.~Nakamura, S.~Uchida, and
  B.~Nachumi,
\newblock Phys. Rev. B {\bf 54}, 7489 (1996).

\bibitem{bianc;prl96}
A.~Bianconi, N.~L. Saini, A.~Lanzara, M.~Missori, T.~Rossetti, H.~Oyanagi,
  H.~Yamaguchi, K.~Oka, and T.~Ito,
\newblock Phys. Rev. Lett. {\bf 76}, 3412 (1996).

\bibitem{bozin;prb99}
E.~S. Bo{\v z}in, S.~J.~L. Billinge, G.~H. Kwei, and H.~Takagi,
\newblock Phys. Rev. B {\bf 59}, 4445 (1999).

\bibitem{radae;prb94i}
P.~G. Radaelli, D.~G. Hinks, A.~W. Mitchell, B.~A. Hunter, J.~L. Wagner,
  B.~Dabrowski, K.~G. Vandervoort, H.~K. Viswanathan, and J.~D. Jorgensen,
\newblock Phys. Rev. B {\bf 49}, 4163 (1994).

\bibitem{gutma;unpub00}
M.~Gutmann, {E. S. Bo\v zin}, and S.~J.~L. Billinge.

\bibitem{billi;ijmpb02}
S.~J.~L. Billinge and P.~M. Duxbury,
\newblock Int. J. Mod. Phys. B  (2002).

\bibitem{billi;prb02}
S.~J.~L. Billinge and P.~M. Duxbury,
\newblock  {\bf 66}, 064529 (2002).

\bibitem{emery;pnas99}
V.~J. Emery, S.~A. Kivelson, and J.~M. Tranquada,
\newblock Proc. Natl. Acad. Sci. USA {\bf 96}, 8814 (1999).

\bibitem{shake;prb95}
H.~Shaked, J.~D. Jorgensen, B.~A. Hunter, R.~L. Hitterman, A.~P. Paulikas, and
  B.~W. Veal,
\newblock Phys. Rev. B {\bf 51}, 547 (1995).

\bibitem{toby;prl90}
B.~H. Toby, T.~Egami, J.~D. Jorgensen, and M.~A. Subramanian,
\newblock Phys. Rev. Lett. {\bf 64}, 2414 (1990).

\bibitem{yamam;prb90}
A.~Yamamoto, M.~Onoda, E.~Takayama–Muromachi, F.~Izumi, T.~Ishigaki, and
  H.~Asano,
\newblock Phys. Rev. B {\bf 42}, 4228 (1990).

\end{thebibliography}
\end{document}